\documentclass[showpacs,showkeys,twocolumn,amsmath,amssymb]{revtex4}
\usepackage{times}
\usepackage{graphicx}
\usepackage{epsfig}
\usepackage{dcolumn}
\usepackage{bm}

\begin{document}
\title{Quantum interference in timed Dicke basis and its effect on bipartite entanglement}
\author{Eyob A. Sete}
\email{eyobas@physics.tamu.edu}
\author{Sumanta Das}
\email{dsumanta31@yahoo.com}
\affiliation{Institute for Quantum Science and
Engineering and Department of Physics and Astronomy, Texas A$\&$M
University, College Station, TX 77843-4242 }
\date{\today}
\begin{abstract} %
We analyze the effect of position dependent excitation phase on the
 properties of entanglement between two qubits formed in atomic systems. We show that
the excitation phase induces a vacuum mediated quantum interference
in the system that affects the dynamical behavior of entanglement
between the qubits. It is also found that the quantum interference
leads to a coherent population transfer between the symmetric and
antisymmetric states which can considerably modify the dynamics of
two-qubit entanglement and can even prevent finite time
disentanglement (sudden death) under certain conditions.
\end{abstract}
 \maketitle
\section{Introduction}
Quantum interference (QI), an intriguing consequences of the
superposition principle has led to numerous fascinating phenomena
such as coherent population trapping \cite{Alz76}, lasing without
inversion \cite{Koc88}, electromagnetically induced transparency
\cite{Har90}, and quantum entanglement \cite{Lon08,Pol08,Rai09}. The
application of QI in generation of bipartite entanglement both in
discrete \cite{Mat09, Ber09} and continuous-variable
\cite{Xio05,Ale08} settings has been the focus of current
investigation. Note that bipartite entanglement involving two atoms,
extensively used for implementations of various quantum information
protocols \cite{Nie04, Ben00,Bou97, Rie04,Bar04,Olm09}, is known to
be quite fragile in the face of decoherence \cite{Nie04,Zur91}. In
view of this, in the past few years considerable effort has been
devoted to the study of dynamical aspect of two atom entanglement in
presence of decoherence \cite{Fic02,Tan04,Dio03,Daf03,Dod04,Min05,
Yu04,Fic06, Das09}.  In one such study \cite{Fic02} it was found
that in contrary to the adverse effect of spontaneous emission on
atomic entanglement \cite{Yu04}, cooperative spontaneous emission in
two atom systems can generate entanglement among the atoms. It is
worth mentioning here that the problem of cooperative spontaneous
emission first addressed by Dicke \cite{Dic54} is known to exhibit
several counter-intuitive phenomena in two atom systems \cite{Aga74}
namely, directed spontaneous emission \cite{Scu06}, Lamb shift
\cite{Scu09}, single photon Dicke superradiance \cite{Set10} and
others. In recent times, with the discovery of atom like behavior of
semiconductor quantum dots \cite{Sch97,Xu07,Pet05} and their
utilization towards solid state quantum computing
\cite{Bar95,Sch97,Xu07, Pet05}, we have a new class of systems where
the phenomenon of cooperative spontaneous emission can be of immense
importance from the context of quantum information sciences.

Recently, Ooi \textit{et al}. \cite{Ooi07} studied the effect of position dependant excitation phase on the population dynamics, intensity and spatial and angular correlations for two two-level atoms interacting via their dipoles. The results show that the excitation phase considerably modifies the dynamics of the system. Later, Das \textit{et al}. \cite{Das08} investigated the
effect of the position dependent excitation phase on the
Dicke cooperative emission spectrum. A strong quantum correlation among the atoms was reported in presence of
the excitation phase. This was attributed to a vacuum mediated QI
generated in the two atom system in presence of the position
dependent excitation phase. The result of \cite{Das08} qualitatively
indicates that the spatial variation of the excitation phase can
affect the generation and evolution of entanglement in the system.
It may be added that, such vacuum mediated QI and its effect has been earlier
studied in atomic systems \cite{Aga74,Zhu96,Zho96,Kei99}. Further, a recent work has
predicted how one can use such QI to protect
bipartite entanglement \cite{Das10}. While these earlier works
utilize the quantum interference that comes about due to the
configuration of the atomic system, we are motivated at studying the
effect of quantum interference induced by the position dependent
excitation phase.

To understand the effects of such QI on the two atom entanglement,
we in this paper perform a systematic study of the time evolution of
entanglement measure for two strongly dipole coupled atoms
undergoing a cooperative spontaneous emission. We consider various
initial quantum states in which the two atoms can be prepared and
explore the effects on the dynamical behavior of entanglement that
results as a consequence of the quantum interference. We explicitly
take into account the position dependent excitation of the atoms by
introducing \textit{timed Dicke basis} \cite{Scu06}. It is important
to understand that the entanglement in a two atom system crucially
depends on the cooperative decay rates, the initial conditions, and
the dipole-dipole interactions \cite{Tan04}, all of which gets
modified due to the quantum interference.
It is worth mentioning here that QI arising from position dependent
phase in such timed Dicke basis was explored in a recent study in
context to population dynamics and photon correlation studies in
two atom systems \cite{Ooi07}. We however in this current work are
interested in investigating the effect of such QI on the entanglement
of two atoms. For instance, for the system initially prepared in the
symmetric timed Dicke state, a coherence between the symmetric and
antisymmetric states is dynamically generated as a result of the QI
between the two pathways to the ground state. This coherence leads
to considerably slow decay of entanglement.

The organization of the paper is as follows. In Sec. II we discuss
our model and write down the dynamical equation for our system using
a master equation approach. Then in Sec. III we discuss the
entanglement measure and derive generalized analytical expressions
in the timed Dicke basis for the two atom system. In Sec. IV, we
then consider two initial conditions: for the atoms prepared in pure
states and mixed state and show explicitly that the vacuum mediated
QI induced by excitation phase can lead to considerable modification
of two atom entanglement behavior. We provide analytical and
numerical results in support of our propositions. Finally, we
summarize our results in Sec. V.

\begin{figure*}[t]
\includegraphics{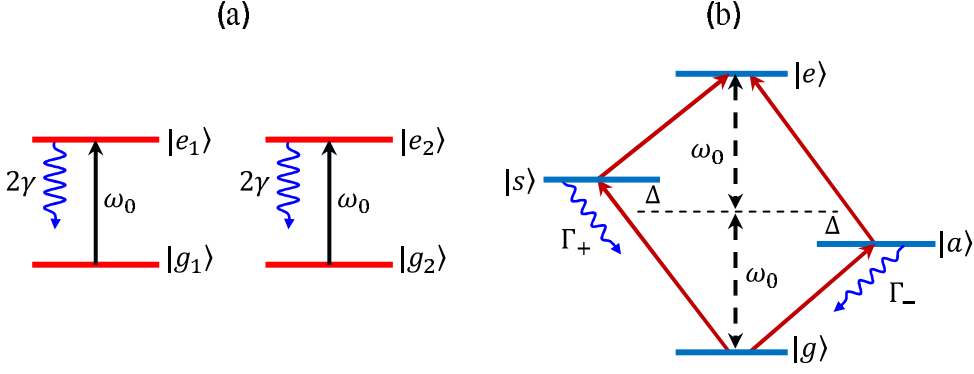}
\caption{(Color Online)Energy level diagram for two two-level atoms in bare basis
(a) and in the timed Dicke basis (b). The frequency shift
$\Delta=\Omega_{12}\cos\varphi$ occurs as a result of dipole-dipole
interaction between the two atoms. The collective states $|s\rangle$
and $|a\rangle$ decays at a rate of $\Gamma_{+}$ and $\Gamma_{-}$
respectively, where $\Gamma_{\pm}=2(\gamma\pm
\gamma_{12}\cos\varphi)$.} \label{fig1}
\end{figure*}

\section{Model and equations of evolution}
We consider a system of two qubits formed by the excited states
$|e_{i}\rangle$ and ground states $|g_{i}\rangle$ ($i=1,2)$ of two
identical two level atoms. The qubits are fixed at positions $r_{1}$
and $r_{2}$ and the inter-atomic distance is less than the
wavelength of the radiation field, $\lambda$. We further assume that
the qubits are coupled to one another by a dipole-dipole interaction
and are coupled to the environment via an interaction with a common
vacuum reservoir. The time evolution of the density operator for
such a two-qubit system can be treated in a master equation
framework and is given by \cite{Aga74}
\begin{align}\label{1}
    \frac{d}{dt}\rho&=
    -i\omega_{0}\sum_{i=1}^{2}[\sigma_{i}^{z},\rho]-i\sum_{i\neq
    j}^{2}\Omega_{ij}[\sigma_{i}^{\dagger}\sigma_{j},\rho]\notag\\
    &-\sum_{i,j=1}^{2}\gamma_{ij}(\rho\sigma_{i}^{\dagger}\sigma_{j}+\sigma_{i}^{\dagger}\sigma_{j}\rho-2\sigma_{j}\rho\sigma_{i}^{\dagger}),
\end{align}
where $\omega_{0}$ is the atomic transition frequency,
$\sigma_{i}^{z}=(\sigma_{i}^{\dagger}\sigma_{i}-\sigma_{i}\sigma_{i}^{\dagger})/2$
is the energy operator with $\sigma^{\dagger}_{i}(\sigma_{i})$ being
the raising (lowering) operator for $i\text{th}$ atom, $\Omega_{ij}$
and $\gamma_{ij}$ for $i\neq j$ are respectively, the dipole-dipole
interaction  term and the cooperative decay rate given by
\begin{align}\label{2}
\Omega_{ij}&=\frac{3}{2}\gamma\bigg[
(1-3\cos^2\theta)\left(\frac{\sin
(k_{0}r_{ij})}{(k_{0}r_{ij})^2}+\frac{\cos(k_{0}r_{ij})}{(k_{0}r_{ij})^3}\right)
\notag\\
&-(1-\cos ^2\theta)\frac{\sin (k_{0}r_{ij})}{k_{0}r_{ij}}\bigg]
\end{align}
and
\begin{align}\label{3}
    \gamma_{ij}&=\frac{3}{2}\gamma\bigg[(1-\cos
^2\theta)\frac{\sin
(k_{0}r_{ij})}{k_{0}r_{ij}}\notag\\
&+(1-3\cos^2\theta)\left(\frac{\cos
(k_{0}r_{ij})}{(k_{0}r_{ij})^2}-\frac{\sin(k_{0}r_{ij})}{(k_{0}r_{ij})^3}\right)\bigg],
\end{align}
where $2\gamma\equiv
2\gamma_{11}=2\gamma_{22}=2|\vec{\wp}_{eg}|^2\omega^3/3\pi\varepsilon_{0}\hbar
c^3$ is the spontaneous decay rate of the individual qubits.
$\vec{\wp}_{eg}$ is the dipole moment, $k_{0}=2\pi/\lambda$ with
$\lambda$ being the wavelength of the emitted radiation  and
$\theta$ is the angle between the direction of the dipole moment and
the line joining the $i\text{th}$ and the $j\text{th}$ qubits, and
$r_{ij}=|\textbf{r}_{i}-\textbf{r}_{j}|$ is the interqubit
distance. In this paper we assume that the orientation of the dipole
moment is random and hence Eqs. \eqref{2} and \eqref{3} simplifies
considerably and take the form

\begin{eqnarray}
 \Omega_{ij} &=& -\gamma\cos (k_{0}r_{ij})/k_{0}r_{ij}, \\
  \gamma_{ij} &=& \gamma \sin (k_{0}r_{ij})/k_{0}r_{ij}.
\end{eqnarray}

We next consider the preparation of initial state of the qubits. For this purpose
we assume that the qubits interacts with a very weak laser field (almost at a single photon level) propagating with a wave-vector $\textbf{k}_{0}$. The interaction with the weak field can lead to a resonant single photon absorption process. It is important to note that we consider the direction of the wave-vector to be different to that of the inter-qubit axis. This thus generate a position dependent excitation phase of the qubits when ever a photon is absorbed.  The excitation process, with the laser field treated classically and in
the rotating wave approximation, can be described by the Hamiltonian
\begin{equation}\label{h}
    V=-\hbar\Omega\sum_{j=1}^{2}(\sigma_{j}^{+}e^{i\textbf{k}_{0}\cdot \textbf{r}_{j}}~e^{-i(\nu_{0}-\omega_{0})t} +\text{adj}.),
\end{equation}
where $\Omega=\textbf{d}_{e_{1}g_{1}}\cdot \mathcal{E
}/\hbar=\textbf{d}_{e_{2}g_{2}}\cdot \mathcal{E }/\hbar$ is the Rabi
frequency and $\nu_{0}$ is the angular frequency of the incident
radiation. Note that the position dependent phase factors in the
Hamiltonian would substantially affect the dynamical behavior of the
correlation in the two qubit system. We except that this in turn
will lead to modification of entanglement among the qubits.  The
investigation of any such modification in the entanglement feature
is the key focus of this paper. In order to investigate the effect
of position dependent excitation phase on the dynamics it proves to
be convenient to work in a basis defined by the phase factors. Such
a basis was introduced in Ref. \cite{Scu06} in context to directed
spontaneous emission from an ensemble of atoms and is also known as
the \textit{timed Dicke basis}. To this end, for our system of two
qubits there are four timed Dicke states:

\begin{eqnarray}
|e\rangle&=& |e_{1}e_{2}\rangle
e^{i\textbf{k}_{0}\cdot\textbf{r}_{1}+i\textbf{k}_{0}\cdot\textbf{r}_{2}}, \\
 |s\rangle &=& \frac{1}{\sqrt{2}}(|e_{1}g_{2}\rangle
e^{i\textbf{k}_{0}\cdot\textbf{r}_{1}}+|g_1e_{2}\rangle
e^{i\textbf{k}_{0}\cdot\textbf{r}_{2}}),\\
 |a\rangle &=& \frac{1}{\sqrt{2}}(|e_{1}g_{2}\rangle
e^{i\textbf{k}_{0}\cdot\textbf{r}_{1}}-|g_1e_{2}\rangle
e^{i\textbf{k}_{0}\cdot\textbf{r}_{2}}),\\
 |g\rangle &=& |g_{1}g_{2}\rangle.
\end{eqnarray}

In terms of this basis the equations of evolution for the elements
of the density operator are:
\begin{subequations}
\begin{equation}\label{4a}
    \dot \rho_{ee}=-4\gamma\rho_{ee},
\end{equation}
\begin{align}\label{4b}
    \dot
    \rho_{es}=&-[3\gamma+\gamma_{12}\cos\varphi+i(\omega_{0}-\Omega_{12}\cos\varphi)]\rho_{es}\notag\\
    &+i\sin\varphi(\gamma_{12}-i\Omega_{12})\rho_{ea},
\end{align}
\begin{align}\label{4c}
    \dot
    \rho_{ea}=&-[3\gamma-\gamma_{12}\cos\varphi+i(\omega_{0}+\Omega_{12}\cos\varphi)]\rho_{ea}\notag\\
    &+i\sin\varphi(\gamma_{12}+i\Omega_{12})\rho_{es},
\end{align}
\begin{equation}\label{4d}
    \dot \rho_{eg}=-2(\gamma+i\omega_{0})\rho_{eg},
\end{equation}
\begin{align}\label{4e}
    \dot
    \rho_{ss}=&-2(\gamma+\gamma_{12}\cos\varphi)\rho_{ss}-i\sin\varphi(\gamma_{12}+i\Omega_{12})\rho_{as}\notag\\
    &+i\sin\varphi(\gamma_{12}-i\Omega_{12})\rho_{sa}+2(\gamma+\gamma_{12}\cos\varphi)\rho_{ee},
\end{align}
\begin{align}\label{4f}
    \dot
    \rho_{aa}=&-2(\gamma-\gamma_{12}\cos\varphi)\rho_{aa}-i\sin\varphi(\gamma_{12}-i\Omega_{12})\rho_{as}\notag\\
    &+i\sin\varphi(\gamma_{12}+i\Omega_{12})\rho_{sa}+2(\gamma-\gamma_{12}\cos\varphi)\rho_{ee},
\end{align}
\begin{align}\label{4g}
    \dot
    \rho_{as}=&-2(\gamma-i\Omega_{12}\cos\varphi)\rho_{as}+i\sin\varphi(\gamma_{12}+i\Omega_{12})\rho_{ss}\notag\\
    &+i\sin\varphi(\gamma_{12}-i\Omega_{12})\rho_{aa}-2i\gamma_{12}\sin\varphi\rho_{ee},
\end{align}
\begin{align}\label{4h}
    \dot
    \rho_{gs}=&-[\gamma+\gamma_{12}\cos\varphi-i(\omega_{0}+\Omega_{12}\cos\varphi)]\rho_{gs}\notag\\
    &+i\sin\varphi(\gamma_{12}-i\Omega_{12})\rho_{ga}\notag\\
    &+2(\gamma+\gamma_{12}\cos\varphi)\rho_{se}+2i\gamma_{12}\sin\varphi\rho_{ae},
\end{align}
\begin{align}\label{4i}
    \dot
    \rho_{ga}=&-[\gamma-\gamma_{12}\cos\varphi-i(\omega_{0}-\Omega_{12}\cos\varphi)])\rho_{ga}\notag\\
    &-i\sin\varphi(\gamma_{12}-i\Omega_{12})\rho_{gs}\notag\\
    &-2(\gamma-\gamma_{12}\cos\varphi)\rho_{ae}+2i\gamma_{12}\sin\varphi\rho_{se},
\end{align}
\begin{align}\label{4j}
    \dot
    \rho_{gg}=&2(\gamma+\gamma_{12}\cos\varphi)\rho_{ss}+2(\gamma-\gamma_{12}\cos\varphi)\rho_{aa}\notag\\
    &+2i\gamma_{12}\sin\varphi(\rho_{as}-\rho_{sa}),
\end{align}
where
$\varphi=\textbf{k}_{0}\cdot(\textbf{r}_{i}-\textbf{r}_{j})=\frac{2\pi}{\lambda}r_{ij}\cos\xi$
with $\xi$ being the angle between the laser propagation direction
and the line joining the two atoms.
\end{subequations}

Inspection of Eqs. \eqref{4b} and \eqref{4c} shows that the presence
of atom-atom interaction gives rise to collective level shift
The level shift arises due to the atom-atom interaction only occurs in states $|s\rangle$ and $|a\rangle$. The other collective states $|e\rangle$ and $|g\rangle$ do not see any level shift due to this interaction as per Eq. (11d). That is, the energy difference between the $|e\rangle$ and $|g\rangle$ remain $2\omega_{0}$. The state $|s\rangle$ is shifted up while the state $|a\rangle$ is shifted down by an equal amount $\Delta=\Omega_{12}\cos \varphi$ from the single photon resonance line as shown in Fig. \ref{fig1}b.
It is interesting to note that one can manipulate the level shift by only orienting the laser field
appropriately with respect to the line joining the two atoms. For
example, $\varphi=\pi/2$, i.e., when the angle between the laser
propagation direction and the line joining the two atoms is $\xi=\pi/3$
and the interatomic distance equal to half of the radiation
wavelength, $r_{12}=\lambda/2$, the level shift vanishes. Thus it is
possible to control the level shift by applying a laser field in a
particular direction without turning off the dipole-dipole
interaction.

Further, we note that the transition probability from the excited
state $|e\rangle$ to the one photon states, $|s\rangle$ and
$|a\rangle$, is the sum of the probability of each transition. Since
it is the probability, and not the probability amplitudes that adds up we don't
expect quantum interference phenomenon to occur. However, the
transition probability from the one photon states, $|s\rangle$ and
$|a\rangle$ to the ground state $|g\rangle$ is obtained by squaring
the sum of the amplitude of each transition. When there is a
coherence between the two states ($|s\rangle$ and $|a\rangle$), this
can lead to quantum interference yielding coherent population
transfer between $|s\rangle$ and $|a\rangle$. Indeed, the
populations in $|s\rangle$ and $|a\rangle$ is coupled to the
coherence $\rho_{as}$ as per Eqs. \eqref{4e}-\eqref{4g}. It is worth to note that this coupling
disappears when the direction of propagation of the laser field is
perpendicular to the interqubit axis $\xi=\pi/2$ ($\varphi=0$).
Therefore, we see that in the presence of a position dependent excitation phase
$\varphi$ quantum interference is induced in the system.
In this paper we hence explore to what extent the quantum interference developed in the
system affects the dynamical properties of the bipartite entanglement between the two
qubits.

\section{Entanglement measure}
In general a state of a quantum system is said to be entangled when
the density operator of the composite system cannot factorize into
that of the individual subsystems. There are several entanglement
measures for two-qubit system in the literature. However, we use the
concurrence, a widely used entanglement monotone, for our purpose.
The concurrence, first introduced by Wooters \cite{Woo98}, is
defined by
\begin{equation}\label{w}
    \mathcal{C}(t)=\text{max}(0,\sqrt{\lambda_{1}}-\sqrt{\lambda_{2}}-\sqrt{\lambda_{3}}-\sqrt{\lambda_{4}}),
\end{equation}
where $\lambda_{1}>\lambda_{2}>\lambda_{3}>\lambda_{4}$.
$\{\lambda{i}\}$ are the eigenvalues of the matrix $\rho\tilde\rho$
in which
$\tilde\rho=\sigma_{y}\otimes\sigma_{y}\rho^{*}\sigma_{y}\otimes\sigma_{y}$
with $\sigma_{y}$ being the Pauli matrix. The concurrence takes
values ranging from 0 to 1. For maximally entangled state
$\mathcal{C}(t)= 1$ and for separable state $\mathcal{C}(t)=0$.

In general, for a dissipative system, without any external driving
field, the density matrix of the qubits system has the form
\begin{equation}\label{dens}
   \rho(t)= \left(
      \begin{array}{cccc}
        \ \rho_{11} & 0 & 0 & \rho_{14} \\
        0& \rho_{22} & \rho_{23} & 0\\
        0& \rho_{32} & \rho_{33} & 0\\
        \rho_{41} & 0 & 0 & \rho_{44} \\
      \end{array}
    \right)
\end{equation}
in the following basis set
\begin{align}\label{basis}
    &|1\rangle=|e_{1}e_{2}\rangle e^{i\textbf{k}_{0}\cdot
    (\textbf{r}_{1}+\textbf{r}_{2})}\notag\\
    &|2\rangle=|e_{1}g_{2}\rangle  e^{i\textbf{k}_{0}\cdot
    \textbf{r}_{1}}\notag\\
    &|3\rangle=|g_{1}e_{2}\rangle e^{i\textbf{k}_{0}\cdot
    \textbf{r}_{2}}\notag\\
    &|4\rangle=|g_{1}g_{2}\rangle.
\end{align}
Note that for a quantum state initially prepared in a block form of
\eqref{dens}, the time-evolved density matrix will have the same
block form, i.e., the zeros remain zero and the nonzero components
evolve in time \cite{Das09,Tan04}.
\begin{figure*}[t]
\includegraphics[width=10cm]{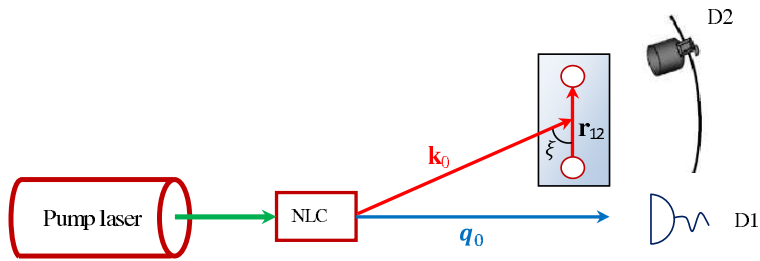}
\caption{(Color online) A scheme illustrating a proposed method to prepare
symmetric timed Dicke state $|s\rangle$. A similar scheme has been
proposed to excite one atom in a cloud of N atoms \cite {Scu06}. The
nonlinear crystal (NLC) down converts the a pump photon into
signal-idler pair. A click on detector (D1) indicates generation of
the pair and hence no click on the second detector (D2)--assuming a
perfect detector--means the other photon ($\textbf{k}_{0}$)
conditionally excite one of the atoms.} \label{fig2}
\end{figure*}
We next proceed to calculate the concurrence for the qubits system
initially prepared in the form of \eqref{dens}. To do so, one has to
determine the matrix $\tilde\rho$ in the basis where $\rho$ is
expressed. Using the definition of the density matrix $\tilde\rho$,
we obtain
\begin{equation}\label{den}
   \tilde\rho(t)= \left(
      \begin{array}{cccc}
        \ \rho_{44} & 0 & 0 & \rho_{14} \\
        0& \rho_{33} & \rho_{23} & 0\\
        0& \rho_{32} & \rho_{22} & 0\\
        \rho_{41} & 0 & 0 & \rho_{11} \\
      \end{array}
    \right).
\end{equation}
Thus the square root of the eigenvalues of the matrix
$\rho\tilde\rho$ are:
\begin{equation}\label{ev}
    \{\sqrt{\lambda_{i}}\}=\{\sqrt{\rho_{22}\rho_{33}}\pm |\rho_{23}|,\sqrt{\rho_{11}\rho_{44}}\pm|\rho_{14}|\}.
\end{equation}
There are two possible expressions for the concurrence, depending on
the values of the eigenvalues. The first case is that when
$|\rho_{23}|+\sqrt{\rho_{22}\rho_{33}}$ be the largest eigenvalue.
This leads to a concurrence
\begin{equation}\label{cu1}
    \mathcal{C}_{1}(t)=2(|\rho_{23}|-\sqrt{\rho_{11}\rho_{44}}).
\end{equation}
While if $|\rho_{14}|+\sqrt{\rho_{11}\rho_{44}}$ is the largest
eigenvalue then the concurrence takes the form
\begin{equation}\label{cu2}
    \mathcal{C}_{2}(t)=2(|\rho_{14}|-\sqrt{\rho_{22}\rho_{33}}).
\end{equation}
Depending on the initial condition used, one of the concurrence
expressions suffices to measure the entanglement present in the
qubits system. Further, inspection of \eqref{cu1} and \eqref{cu2}
shows that $\mathcal{C}_{1}(t)$ would be positive and hence the
measure of entanglement when the one photon coherence is larger the
square root of the product of the populations in the excited and
ground states. On the other hand, for $\mathcal{C}_{2}(t)$ to be a
measure of entanglement for the system the two photon coherence
should be greater than the square root of the product of the
population in one photon excited states.

In order to gain insight into the physics it is convenient to
express the concurrences in terms of timed Dicke basis introduced
earlier. To do so, one has to apply a unitary transformation  $U\rho
U^{\dagger}$ on the density matrix given by \eqref{dens}. The matrix U is given by
\begin{equation}
U=\left(
  \begin{array}{cccc}
    1 & 0 & 0 & 0 \\
    0 & \frac{1}{\sqrt{2}} & -\frac{1}{\sqrt{2}} & 0 \\
    0 & \frac{1}{\sqrt{2}} & \frac{1}{\sqrt{2}} & 0 \\
    0 & 0 & 0& 1 \\
  \end{array}
\right).
\end{equation}
The elements of the density matrix $U\rho U^{\dagger}$ is related to
that of $\rho$ by
\begin{align}\label{tra}
&\rho_{ee} = \rho_{11} \notag\\
& \rho_{eg} = \rho_{14}\notag\\
&\rho_{aa} = \frac{1}{2}(\rho_{22}+\rho_{33}-(\rho_{23}+\rho_{32})) \notag\\
&\rho_{ss}= \frac{1}{2}(\rho_{22}+\rho_{33}+\rho_{23}+\rho_{32}) \notag\\
&\rho_{as} = \frac{1}{2}(\rho_{22}-\rho_{33}+\rho_{23}-\rho_{32}) \notag\\
&\rho_{sa} = \frac{1}{2}(\rho_{22}-\rho_{33}-(\rho_{23}-\rho_{32})).
\end{align}
Therefore the concurrence can be expressed in terms of the timed
Dicke basis as
\begin{equation}\label{cu}
    \mathcal{C}(t)=\text{max}(0,\mathcal{C}_{1}(t),\mathcal{C}_{2}(t)),
\end{equation}
where
\begin{equation}\label{cu3}
    \mathcal{C}_{1}(t)=\sqrt{(\rho_{ss}-\rho_{aa})^2+4[\text{Im}(\rho_{as})]^2}-2\sqrt{\rho_{ee}\rho_{gg}}
\end{equation}
\begin{equation}\label{cu4}
    \mathcal{C}_{2}(t)=2|\rho_{eg}|-\sqrt{(\rho_{ss}+\rho_{aa})^2+4[\text{Re}(\rho_{as})]^2}.
\end{equation}
This expression for concurrence will be used in the following
section to study the dynamical evolution of entanglement in the
two-qubit system by considering various initial conditions.

\section{Entanglement dynamics of two identical qubits}

Using the entanglement measure introduced in the previous section we
study the effect of the position dependent excitation phase by
considering pure and mixed state as initial conditions.

\subsection{Initial pure states}
In the two-qubit system one might consider a pure separable or
entangled state as an initial condition. For instance, for pure
separable state, one can take the two atom excited state,
$|e\rangle$. Even though this is unentangled state at the initial
time, the interaction of the atoms with the environment leads to
weak transient entanglement \cite{Fic02,Tan04}. The effect of
quantum interference induced by position dependent excitation phase
is unimportant in this case and thus we rather focus on pure
entangled state as an initial condition.

We take the initial state of the two-qubit system to be the
symmetric state $|s\rangle$. This state is a pure maximally
entangled state and can be prepared using correlated pair of photons
generated from a parametric down conversion process in which one of
the photons is sent to a detector (D1) and the other is directed
towards the atoms. A click on the detector (D1) tells us that the
other photon is sent to the atoms. If the second detector (D2)
registrars a count then no atom is excited. However, if D1 shows a
click and D2 does not then we know that one of the atoms is excited,
but we don't know which one (see Fig. \ref{fig2}). This leads to a
superposition state $|s\rangle$. Recently, Thiel \textit{et al}.
\cite{Thi07} proposed a method to prepare all the symmetric states
using a linear optical tools. We seek to investigate the dynamics of
entanglement of the two qubits as they interact with the
environment and with each other via their electric-dipoles.

In terms of the timed Dicke basis the initial density matrix has
only one nonzero element namely $\rho_{ss}(0)=1$; all the rest of
matrix elements are zero. Since there is no initial two photon
coherence $\rho_{eg}(0)=0$, according to Eq. \eqref{4d}, it remains
zero all the time. As a consequence the expression given by
\eqref{cu4} will be negative and hence cannot be used as
entanglement measure. Moreover, it is easy to see that for initial
condition we considered,  $\rho_{ee}(t)=0$. In view of this
expression \eqref{cu3} takes the form
\begin{equation}\label{cu5}
    \mathcal{C}_{1}(t)=\sqrt{(\rho_{ss}-\rho_{aa})^2+4[\text{Im}(\rho_{as})]^2}>0
\end{equation}
and thus the concurrence can be written as
\begin{equation}\label{cu6}
    \mathcal{C}(t)=\text{max}(0,\mathcal{C}_{1}(t)).
\end{equation}
This expression shows that the concurrence is unity at $t=0$ as it
should be.

Disregarding the relative phase shift ($\varphi=0$) the solutions of
 the elements of the density matrix in Eq. \eqref{cu5} turn
out to be $\rho_{ss}(t)=\exp[{-2(\gamma+\gamma_{12})t}]$,
$\rho_{aa}(t)=0$, and $\rho_{as}=0$, which leads to
 \begin{equation}\label{cu6}
    \mathcal{C}(t)=\text{max}(0,e^{-2(\gamma+\gamma_{12})t}).
 \end{equation}
We immediately see that the concurrence depends only on the
symmetric state population, $\rho_{ss}(t)$. As there is no single
photon coherence generated in this case, population transfer between
the $|s\rangle$ and $|a\rangle$ does not occur. As a result the
initial entanglement experiences an enhanced decay due to the
collective decay rate ($\gamma_{12}$) and goes asymptotically to
zero as $t\rightarrow \infty$. For nonidentical atoms, however, even
though the entanglement has the same behavior as identical atoms at
the initial time, it exhibits revival at later times \cite{Fic02}.
Here the detuning plays an important role in creating coherence
between the symmetric and antisymmetric states, which is the basis
for entanglement in the two-qubit system. In the following we rather
show, by taking into account the spatial phase dependence of the
atomic states, that quantum interference in the system leads to a
population transfer between the symmetric and antisymmetric states
and hence generation of coherence ($\rho_{as}$).
\begin{figure}[b]
\includegraphics[width=8cm]{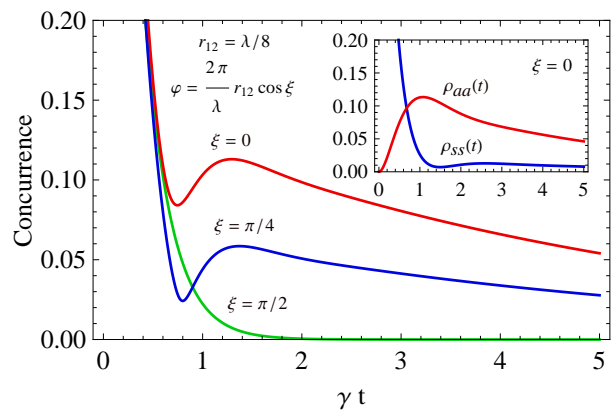}
\caption{(Color online) Plots of the time evolution of concurrence
$\mathcal{C}_{1}(t)$ with initial condition $\rho_{\text{ss}}(0)=1$, for
interatomic distance $r_{12}=\lambda/8$ ($\gamma_{12}/\gamma=0.9,
\Omega_{12}/\gamma=-0.9$) for different values of $\xi$-the angle
between the direction of propagation of the laser and the line
joining the two atoms. The inset shows the populations in the
symmetric and antisymmetric states for $\xi=0 (\varphi=\pi/4)$. }
\label{sfig2}
\end{figure}
\begin{figure}[t]
\includegraphics[width=8cm]{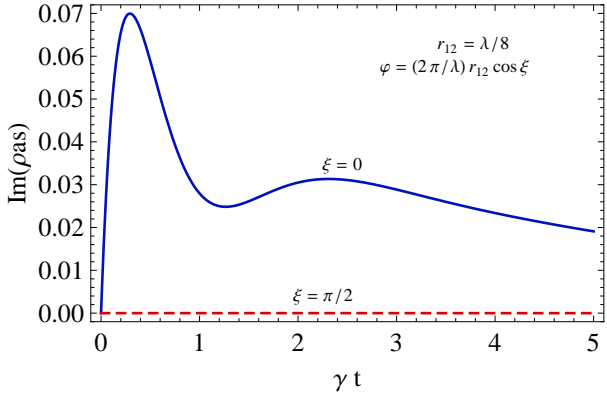}
\caption{(Color online) Plots the imaginary part of coherence between the symmetric
and antisymmetric states $\rho_{\text{as}}$ with initial condition
$\rho_{\text{ss}}(0)=1$, for interatomic distance $r_{12}=\lambda/8$
($\gamma_{12}/\gamma=0.9, \Omega_{12}/\gamma=-0.9$) and for
$\xi=0$.} \label{sfig3}
\end{figure}

The phase shift that an atom experiences during the excitation
process contain physical information about the excited atom. For
example, in the phase factor associated with an excited atom
$\exp(i\textbf{k}\cdot \textbf{r}_{j})$ the term $\textbf{k}\cdot
\textbf{r}_{j}=\omega_{0}\hat n\cdot \textbf{r}_{j}/c\equiv
\omega_{0}t_{i}$ indicates that the atom located at position $r_{j}$
is excited at different times $t_{i}$. This has been discussed in
the context of directed spontaneous emission and collective Lamb
shift in recent years \cite{Scu06,Scu09}. Here we present how this
phase factor can be used to improve the entanglement at later times.

In one photon subspace [$\rho_{ss}(0)=1$] and for nonzero spatial
excitation phase the important equations read \cite{Das08}
\begin{align}\label{4en}
    \dot
    \rho_{ss}=&-2(\gamma+\gamma_{12}\cos\varphi)\rho_{ss}-i\sin\varphi(\gamma_{12}+i\Omega_{12})\rho_{as}\notag\\
    &+i\sin\varphi(\gamma_{12}-i\Omega_{12})\rho_{sa},
\end{align}
\begin{align}\label{4fn}
    \dot
    \rho_{aa}=&-2(\gamma-\gamma_{12}\cos\varphi)\rho_{aa}-i\sin\varphi(\gamma_{12}-i\Omega_{12})\rho_{as}\notag\\
    &+i\sin\varphi(\gamma_{12}+i\Omega_{12})\rho_{sa},
\end{align}
\begin{align}\label{4gn}
    \dot
    \rho_{as}=&-2(\gamma-i\Omega_{12}\cos\varphi)\rho_{as}+i\sin\varphi(\gamma_{12}+i\Omega_{12})\rho_{ss}\notag\\
    &+i\sin\varphi(\gamma_{12}-i\Omega_{12})\rho_{aa}.
\end{align}
These equations fully describe the dynamical behavior of the
population transfer between the symmetric and antisymmetric states
and the coherence developed between them under the given initial
condition. We particularly note that the coherence critically
depends on the spatial excitation phase ($\varphi$). It is not
difficult to see from these equations that for a laser propagating
perpendicular to the interqubit axis ($\varphi=0$) there will be no
coherence, which in turn implies the initial population in the
symmetric state directly decays to the ground state without ever
being transferred to the antisymmetric state. In this decay process
the maximum entanglement present at the initial time will be washed
out in short time. Therefore, for this particular initial condition,
one has to play around with the spatial excitation phase to avoid
enhanced decay of the entanglement.

Using the analytical solutions of the Eqs. \eqref{4en}-\eqref{4gn}
the concurrence can be expressed as
\begin{equation}\label{cu77}
    \mathcal{C}(t)=\text{max}(0,\mathcal{C}_{1}(t)),
\end{equation}
where
\begin{align}\label{cu8}
    \mathcal{C}_{1}(t)=&e^{-2\gamma t}[(\cos\varphi\cosh 2\gamma_{12}t-\sinh 2\gamma_{12}t)^2\notag\\
    &+\sin^2\varphi\cos^2
    2\Omega_{12}t]^{1/2}.
\end{align}
Inspection of \eqref{cu8} shows that the presence of the excitation
phase brings in the dipole-dipole interaction ($\Omega_{12}$) into
the dynamics. This is in contrast with the case where $\varphi=0$ in
which the concurrence is independent of the interatomic interaction.
Note that it is the initial preparation of the state that determines
the dynamical behavior of the two-qubit system. To better understand
to what extent the excitation phase modifies the concurrence and
hence the entanglement between the qubits, we graphically present
the concurrence in Fig. \ref{sfig2}.

In Fig. \ref{sfig2} we show the evolution of the concurrence as a
function of the angle between the direction of propagation of the
laser and the line joining the two atoms ($\xi$) for the two-qubit
system prepared initially in the symmetric state $|s\rangle$ and for
interatomic distance $r_{12}=\lambda/8$. As we have discussed
earlier the concurrence corresponding to $\varphi=0$ exhibits a
sharp decrease and ultimately goes to zero for $t\rightarrow
\infty$. The situation for nonzero excitation phase is different;
the concurrence sharply diminishes during the decay time of the
symmetric state $[2\gamma+\gamma_{12}\cos\varphi]^{-1}$ and shows a
bit of revival and decays slowly before it goes to zero at
$t\rightarrow\infty$. This can be understood by looking at the inset
of the Fig. \ref{sfig2}, where we plotted the time evolution of
populations in the symmetric and antisymmetric states. As can be
clearly seen from this inset that  for $\varphi\neq 0 (\xi=0)$, quantum
interference leads to coherent transfer of population from the
initially populated state $|s\rangle$ to antisymmetric state
$|a\rangle$ \cite{Ooi07} and  hence generation of coherence between these levels
as illustrated in Fig. \ref{sfig3}. This coherence is responsible
for the entanglement observed between the qubits.

\subsection{Initial mixed state}
We next consider the two qubits initially prepared in a mixed entangled state
\cite{Yu04} given by the density matrix
\begin{equation}\label{in}
    \rho(0)=\frac{1}{3}(a|1\rangle\langle 1|+(1-a)|4\rangle\langle 4|+(b+c)|\Phi\rangle \langle \Phi|)
\end{equation}
in which
$|\Phi\rangle=\frac{1}{\sqrt{b+c}}(\sqrt{b}|2\rangle+e^{i\chi}\sqrt{c}|3\rangle)$
and the normalization condition reads $(1+b+c)/3=1$. Here $a, b, c$
and $\chi$ are independent parameters which determine the initial
state of the two entangled qubits. Note that the above state is a form of
generalized Werner state. The initial condition given by
\eqref{in} can be written in the basis of \eqref{dens} as
\begin{equation}\label{ro}
    \rho(0)=\frac{1}{3}\left(
              \begin{array}{cccc}
                a & 0 & 0 & 0 \\
                0 & b & z & 0 \\
                0 & z^{*} & c&0 \\
                0 & 0 & 0 & 1-a \\
              \end{array}
            \right)
\end{equation}
where $z=\sqrt{bc}~e^{i\chi}$ is some initial single photon coherence in the system
and $\chi$ is the respective phase of the coherence.
Now applying the transformation given
by \eqref{tra}, the initial density matrix elements for $b=c=|z|=1$
become
\begin{align*}
   & \rho_{ee}(0)=a/3, \rho_{aa}(0)=(1-\cos \chi)/3,\rho_{gg}(0)=(1-a)/3
\notag\\
&\rho_{ss}(0)=(1+\cos \chi)/3, \rho_{as}(0)=\frac{i}{3}\sin\chi.
\end{align*}
Since $\rho_{es}(0)=\rho_{ea}(0)=\rho_{gs}(0)=\rho_{ga}(0)=0$, the
form of the initial density matrix remain the same, i.e., all the
zero elements remain zero and the all the rest evolves in time.
Under this scenario the expression given by \eqref{cu4} will be
negative and hence cannot be an entanglement measure for the qubit
system. Therefore, \eqref{cu3} is the only candidate left to
quantify the entanglement between the qubits.
\begin{figure}[t]
\includegraphics{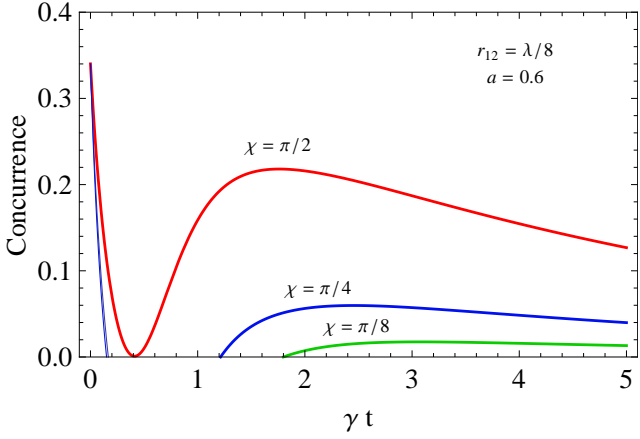}
\caption{(Color online)Plots of the time evolution of concurrence $\mathcal{C}(t)$
with initial condition $b=c=|z|=1$ and for $a=0.6$,
$r_{12}=\lambda/8$, and for different values of the initial phase
$\chi$} \label{sfig4b}
\end{figure}
For $\varphi=0$ the system of equations governing the dynamics of
the two qubits can be solved analytically. Using these solutions,
solved under the initial condition \eqref{ro}, the expression that
describes the entanglement between the qubits, $\mathcal{C}_{1}(t)$,
turns out to be
\begin{align}\label{c6}
  \mathcal{C}_{1}(t)&=\frac{2}{3}e^{-2\gamma t}\big\{\big[\left(\cos\chi\cosh 2\gamma_{12}t-\sinh
  2\gamma_{12}t+a\eta_{1}(t)\right)^2\notag\\
  &+\sin^2 \chi \cosh
  2\Omega_{12}t\big]^{1/2}-\sqrt{3a(1-\eta_{2}(t))}\big\},
\end{align}
where
\begin{align}\label{eta1}
    \eta_{1}(t)&=\frac{(\gamma^2+\gamma_{12}^2)}{\gamma_{12}^2-\gamma^2}\sinh 2\gamma_{12}t\notag\\
    &+\frac{2\gamma\gamma_{12}}{\gamma_{12}^2-\gamma^2}(e^{-2\gamma t}-\cosh
    2\gamma_{12}t),
\end{align}
\begin{align}\label{rgg}
    \eta_{2}(t)&=\frac{a}{3}e^{-4\gamma t}+\frac{2}{3}e^{-2\gamma t}\big[\cosh
    2\gamma_{12} t-\cos\chi\sinh 2\gamma_{12}t \notag\\
    &+a\frac{(\gamma^2+\gamma_{12}^2)}{\gamma_{12}^2-\gamma^2}(e^{-2\gamma t}-\cosh 2\gamma_{12} t)
    \notag\\
    &-a\frac{2\gamma\gamma_{12}}{\gamma_{12}^2-\gamma^2}\sinh
    2\gamma_{12}t)\big].
\end{align}
We immediately see from this result that the concurrence depends on
the parameters $a$, which characterizes the initial populations of
the doubly excited state \textit{i.e.} when both the qubits are
excited and on the phase parameter $\chi$ which determines the
initial populations in the symmetric and antisymmetric states as
well as the coherence between them. If we consider that the qubits
are coupled independently to their respective vacuum environment
($\gamma_{12}=0$) and are well separated in position ($r_{12} \gg
\lambda$) so that the dipole-dipole interaction
($\Omega_{12}\rightarrow 0$), $\mathcal{C}_{1}(t)$ reduces to
\begin{align*}
    \mathcal{C}_{1}(t)=\frac{2}{3}e^{-2\gamma t}\Big[1-\sqrt{a(1-a+2\alpha^2+\alpha^4a)}\Big]
    \end{align*}
where $\alpha(t)=\sqrt{1-\exp(-2\gamma t)}$. Note that
$\mathcal{C}_{1}(t)$ is independent of the initial phase $\chi$.
This coincides with the earlier results of Yu and Eberly
\cite{Yu04}.

\begin{figure}[t]
\includegraphics{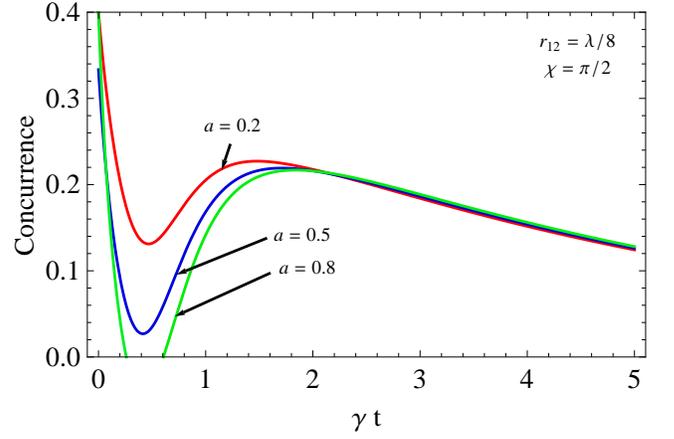}
\caption{(Color online)Plots of the time evolution of concurrence with initial
condition $b=c=|z|=1$ and for $\chi=\pi/2$, $r_{12}=\lambda/8$
($\gamma_{12}/\gamma=0.9, \Omega_{12}/\gamma=-0.9$), for different
values of the initial populations $a$.} \label{sfig4c}
\end{figure}

In the following we study the dependence of the concurrence and
hence the entanglement between the qubits on various system
parameters. Figure \ref{sfig4b} shows the time evolution of the
concurrence for $r_{12}=\lambda/8$ and $a=0.6$ and for different
values of the initial phase, $\chi$. We observe from this figure
that the initial entanglement between the qubits vanishes and
exhibits revival. The amplitude of revival and the revival time
(the time at which the entanglement revive in the system) are directly
related to the initial coherence in the system. The higher the
initial coherence the higher the amplitude of revival and the
shorter the revival time is. Not surprisingly the magnitude of
revival diminishes when the initial coherence decreases. Now keeping
the initial coherence at its maximum value ($\chi=\pi/2)$, we
investigate the influence of the population distribution between the
excited and ground states on the concurrence. Figure \ref{sfig4c}
shows the evolution of the concurrence for $\chi=\pi/2$ and for
different values of $a$. This figure indicates that when the initial
population in the excited state grows the transient entanglement
falls sharply and even disappears for $a=0.8$
($\rho_{ee}(0)\approx0.27$) in the short time window. The
entanglement then shows revival and a slowly damping behavior
afterwards for all values of initial populations.
\begin{figure}[t]
\includegraphics{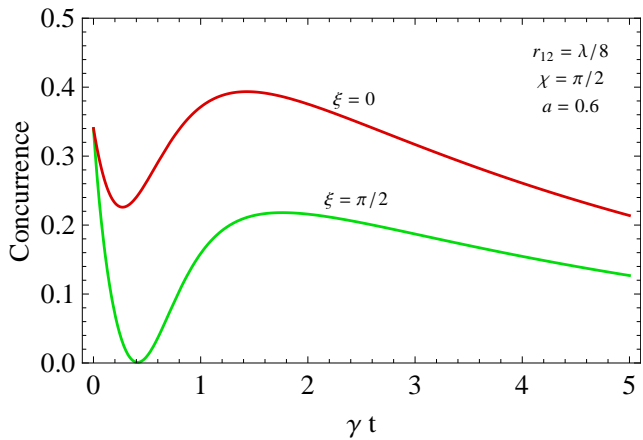}
\caption{(Color online)Plots of the evolution of concurrence with initial
condition $b=c=|z|=1$ and for $r_{12}=\lambda/8$
($\gamma_{12}/\gamma=0.9, \Omega_{12}/\gamma=-0.9$), $\chi=\pi/2$
and for the direction of propagation of the laser field
perpendicular ($\xi=\pi/2$) and parallel ($\xi=0$) to the line
joining the two atoms.} \label{sfig4}
\end{figure}
\begin{figure}[t]
\includegraphics{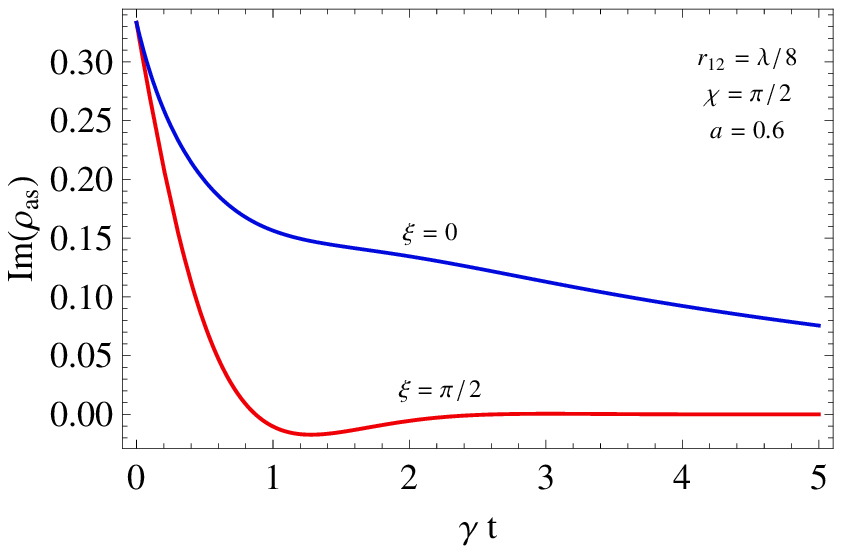}
\caption{(Color online)Plots of the imaginary part of the one photon coherence
$\rho_{\text{as}}$ with initial condition $b=c=|z|=1$ and for
$r_{12}=\lambda/8$ ($\gamma_{12}/\gamma=0.9,
\Omega_{12}/\gamma=-0.9$), $\chi=\pi/2$ and for the direction of
propagation of the laser field perpendicular ($\xi=\pi/2$) and
parallel ($\xi=0$) to the line joining the two atoms.} \label{sfig5}
\end{figure}

We next analyze the evolution of entanglement in the system by
introducing the spatial excitation phase $\varphi$ into the
dynamics. By comparing the previous results for $\varphi=0$ with the
numerical plots for $\varphi\neq 0$, we discuss the effect of the
excitation phase on the entanglement dynamics. Our results are
summarized in Figs. \ref{sfig4} and \ref{sfig5}. In Fig.
\ref{sfig4}, we present a comparison of concurrence taking into
account the spatial excitation phase $\varphi=\pi/4$ ($\xi=0$) and
in the absence of excitation phase, $\varphi=0$ ($\xi=\pi/2$) for
interatomic distance less than the radiation wavelength,
$r_{12}=\lambda/8$. Recall that
$\varphi=(2\pi/\lambda)r_{12}\cos\xi$, where $\xi$ is the angle
between the laser propagation direction and the line joining the two
atoms. These plots clearly show that the excitation phase
effectively protects the initial entanglement from experiencing a
sudden death and even enhances the entanglement from its initial
value during the revival period. The amount of entanglement then
drops gradually and approaches zero at $t\rightarrow \infty$. It is
worth noting that the spatial excitation phase creates additional
coherence and hence improves the revival magnitude over that
observed for the case $\varphi=0$. This enhanced coherence, as shown
in Fig. \ref{sfig5}, is a signature of stronger entanglement between the qubits

\section{conclusion}
We have investigated the effect of quantum interference induced by
position dependent excitation phase on the dynamical behavior of
entanglement between two interacting qubits coupled to a common
vacuum reservoir. We have considered both pure and mixed entangled
states for our analysis. Our results show that for the atoms
initially prepared in a symmetric state, the excitation phase
induces quantum interference in the two-qubit system that leads to
coherent population transfer between the symmetric and antisymmetric
states. This thus creates a coherence which in effect slows down the
otherwise fast decay of two-qubit entanglement considerably. Hence
we find that the evolution of entanglement crucially depends on the
coherence between the symmetric and antisymmetric states.
Furthermore, when the qubits are prepared in a Werner type mixed
entangled state the entanglement is known to suffer sudden death.
However, if one takes into account the excitation phase into the
dynamics the entanglement exhibits revival. This revival is
attributed to strong coherence developed between the symmetric and
antisymmetric states. A viable candidate for realization of our
findings would be semiconductor quantum dots. Note that coupled
quantum dots with interdot distance less than the radiation
wavelength has already been investigated in context to quantum gates
\cite{Sti06} and photoluminescence spectra \cite{Sch97}. As a future
perspective, one can further study the effect of virtual processes
on the dynamics of the system as these processes are known to
influence the evolution of the symmetric and antisymmetric states.

\begin{acknowledgements}
We thank Marlan O. Scully for helpful discussions and gratefully
acknowledge support from the NSF Grant No. EEC-0540832 (MIRTHE ERC),
the Office of Naval Research, and the Robert A. Welch Foundation
(A-1261). E. A. S. is supported by the Herman F. Heep
and Minnie Belle Heep Texas A$\&$M University Endowed Fund
held and administered by the Texas A$\&$M Foundation.
\end{acknowledgements}


\begin{thebibliography}{99}

\bibitem{Alz76}
G. Alzetta, A. Gozzini, L. Moi, and G. Orriols, Nuovo Cimento {\bf 36B}, 5 (1976);
E. Arimondo in Progress in Optics ed. E. Wolf, {\bf 35}, 257 (1996).

\bibitem{Koc88} O. Kocharovskaya and Ya. I. Khanin, JETP Lett. \textbf{48}, 581
(1988); S. E. Harris, Phys. Rev. Lett. \textbf{62}, 1033 (1989); M.
O. Scully, S. Y. Zhu, and A. Gavrielides, Phys. Rev. Lett.
\textbf{62}, 2813 (1989).

\bibitem{Har90}
S. E. Harris, J. E. Field, and A. Imamoglu, Phys. Rev. Lett. {\bf 64}, 1107 (1990);
K. -J. Boller, A. Imamoglu, and S. E. Harris, Phys. Rev. Lett. {\bf 66},  2593 (1991);
M. Fleischhauer, A. Imamoglu, and J. P. Marangos, Rev. Mod. Phys. {\bf 77}, 633 (2005).

\bibitem{Lon08} S. Longhi, Phys. Rev. Lett. {\bf 101}, 193902
(2008); S. Longhi, Phys. Rev. B {\bf 79}, 245108 (2009).

\bibitem{Pol08}  A. Politi, M. J. Cryan, J. G. Rarity, S. Yu, and J. L. O'Brien,
 Science {\bf320}, 646 (2008).


\bibitem{Rai09} A. Rai and G. S. Agarwal, Phys. Rev. A {\bf 79}, 053849 (2009).


\bibitem{Mat09} J. C. F. Matthews, A. Politi, A. Stefanov, and J. L. O'Brien, Nature Photonics {\bf 3}, 346  (2009).

\bibitem{Ber09} D. W. Berry and H. M. Wiseman, Nature Photonics {\bf 3}, 317
  (2009).

\bibitem{Xio05} H. Xiong, M. O. Scully, and M. S. Zubairy, Phys. Rev.
Lett. {\bf 94}, 023601 (2005); E. Alebachew, Phys. Rev. A {\bf 76},
023808 (2007); E. Alebachew, Opt. Commun. {\bf 280}, 133 (2007)

\bibitem{Ale08} E. A. Sete, Opt. Commun. {\bf 281}, 6124 (2008); A. Rai, S. Das, and G. S. Agarwal, Optics Express {\bf 18},
6241 (2010).

\bibitem{Nie04}
M. Nielsen, and I. Chuang, \textit{Quantum Computation and Quantum
Information} (Cambridge Univ. Press, Cambridge 2004).

\bibitem{Ben00}C.H. Bennett \textit{ et al}., Phys. Rev. Lett. \textbf{70}, 1895
(1993); C. H. Bennet and D. P. DiVicenzo, Nature \textbf{404}, 247
(2000); D. P. DiVincenzo, Science {\bf 270}, 255 (1995)

\bibitem{Bou97} D. Bouwmeester \textit{ et al}.  Nature \textbf{390},
375 (1997).

\bibitem{Rie04} M. Riebe \textit{ et al}., Nature  \textbf{429}, 734 (2004).

\bibitem{Bar04} M. D. Barrett  \textit{ et al}., Nature  \textbf{429}, 737 (2004).

\bibitem{Olm09} S. Olmschenk \textit{ et al}., Science  \textbf{323}, 486 (2009).

\bibitem{Zur91}
W. H. Zurek, Physics Today, 36 Oct. (1991); W. H. Zurek, Rev. Mod.
Phys. {\bf 75}, 715 (2003).

\bibitem{Fic02} Z. Ficek and S. Swain , J. Mod. Opt {\bf 49}, 3
(2002).

\bibitem{Dio03} L. Diosi,  \textit{Irreversible Quantum Dynamics} eds. F. Benatti and
R. Floreanini (New York: Springer, 2003).

\bibitem{Daf03} S. Daffer, K. Wodkiewicz,
and J. K. McIver, Phys. Rev. A \textbf{67}, 062312 (2003).

\bibitem{Tan04}R. Tanas and Z. Ficek , J. Opt. B {\bf 6}, S90
(2004).

\bibitem{Dod04} P. J. Dodd
and J. J. Halliwell, Phys. Rev. A \textbf{69}, 052105 (2004).

\bibitem{Min05} F. Mintert,
A. R. R. Carvalho, M. Kus, and A. Buchleitner, Phys. Rep.
\textbf{415}, 207 (2005); J. von Zanthier, T. Bastin, and G.S.
Agarwal, Phys. Rev. A {\bf 74}, 061802(R) (2006).

\bibitem{Yu04}
T. Yu, and J. H. Eberly, Phys. Rev. Lett. {\bf 93}, 140404 (2004);
M. P. Almeida \textit{et. al.}, Science {\bf 316}, 579 (2007); J.
Laurat, K. S. Choi, H. Deng, C.W. Chou, and H.J. Kimble, Phys. Rev.
Lett. {\bf 99}, 180504 (2007).

\bibitem{Fic06}
Z. Ficek and R. Tanas, Phys. Rev. A {\bf 74}, 024304 (2006);
\textit{ibid} {\bf 77}, 054301 (2008).

\bibitem{Das09}
S. Das and G. S. Agarwal, J. Phys. B. {\bf 42}, 205502 (2009); M.
Ban, Eur. Phys. J. D \textbf{58}, 415 (2010).

\bibitem{Dic54}
R. H. Dicke , Phys. Rev. {\bf 93}, 99 (1954).

\bibitem{Aga74} G. S. Agarwal, Springer Tracts in Modern Physics: Quantum
Optics (Springer-Verlag, Berlin, 1974).

\bibitem{Scu06} M. O. Scully, E.S. Fry, C.H.Raymond Ooi, and K. Wodkiewicz, Phys. Rev. Lett. \textbf{96}, 010501
(2006); M. O. Scully, Laser Phys. \textbf{17}, 635 (2007).

\bibitem{Scu09} M. O. Scully, Phys. Rev. Lett. \textbf{102}, 143601
(2009).

\bibitem{Set10} E. A. Sete, A. A. Svidzinsky, H. Eleuch, Z. Yang, R.D.
Nevels, and M.O. Scully, J. Mod. Opt. \textbf{57}, 1311 (2010) and
references therein.

\bibitem{Roh10} R. Röhlsberger, K. Schlage, B. Sahoo, S. Couet, and R. Rüffer, Science \textbf{328}, 1248 (2010).

\bibitem{Sch97} G.
Schedelbeck \textit{et al.}, Science {\bf 278}, 1792 (1997); M.
Bayer \textit{et al.}, Science {\bf 291}, 451 (2001); H. J. Krenner,
\textit{et al.} Phys. Rev. Lett. {\bf 94}, 057402 (2005).

\bibitem{Xu07} X. Xu \textit{et. al.}, Science {\bf 317}, 929 (2007).

\bibitem{Pet05} J. R. Petta \textit{et al.}, Science {\bf 309}, 2180
(2005).

\bibitem{Bar95}
A. Barenco, D. Deutsch, A. Ekert, and R. Jozsa, Phys. Rev. Lett.
{\bf 74}, 4083 (1995).

\bibitem{Ooi07} C. H. Raymond Ooi, B.-G. Kim, and H.-W. Lee, Phys. Rev.
A. \textbf{75}, 063801 (2007).

\bibitem{Das08} S. Das, G. S. Agarwal, and M. O. Scully, Phys. Rev.
Lett. \textbf{101}, 153601 (2008).

\bibitem{Zhu96}
S.-Y. Zhu and M. O. Scully, Phys. Rev. Lett. {\bf 76}, 388 (1996); M. O. Scully and S.-Y. Zhu, \textit{Science} {\bf 281}, 1973 (1998).

\bibitem{Zho96}
P. Zhou and S. Swain, Phys. Rev. Lett. {\bf 77}, 3995 (1996).

\bibitem{Kei99}
C. H. Keitel, Phys. Rev. Lett. {\bf 83}, 1307 (1999).

\bibitem{Das10}
S. Das and G. S. Agarwal,  Phys. Rev. A {\bf 81}, 052341 (2010)

\bibitem{Woo98} W. K. Wootters, Phys. Rev. Lett. \textbf{80}, 2245
(1998).

\bibitem{Thi07} C. Thiel, J. von Zanthier, T. Bastin, E. Solano, and G. S.
Agarwal, Phys. Rev. Lett. \textbf{99}, 193602 (2007).
\bibitem{Sti06} E.A. Stinaff \textit{et. al.}, Science \textbf{311}, 636
(2006); L. Robledo \textit{et. al.}, Science \textbf{320}, 722
(2008).

\end{thebibliography}
\end{document}